\documentclass[article]{aa}
\usepackage{natbib}
\usepackage{graphicx}
\usepackage{color}
\usepackage{sidecap}
\begin{document}

  \title{Observed and Simulated Power Spectra of Kinetic and Magnetic Energy retrieved with 2D inversions}
   \titlerunning{Power Spectra of Solar Internetwork}
   \author{S.~Danilovic \inst{1} \and M.~Rempel\inst{2} \and M.~van~Noort \inst{1}  \and R.~Cameron \inst{1} }

   \institute{Max-Planck-Institut f\"ur Sonnensystemforschung, Justus-von-Liebig-Weg 3
37077 G\"ottingen, Germany \and 
 High Altitude Observatory, NCAR, P.O. Box 3000, Boulder, Colorado 80307, USA}

%   \date{Received ; accepted}
   \date{\today}

\abstract
 {Information on the origin of internetwork magnetic field is hidden at the smallest spatial scales.}
% aims heading (mandatory)
  {We try to retrieve the power spectra with certainty to the highest spatial frequencies allowed by current instrumentation.}
% methods heading (mandatory)
  {For this, we use 2D inversion code that were able to recover information up to the instrumental diffraction limit.}
% results heading (mandatory)
  {The retrieved power spectra have shallow slopes extending further down to much smaller scales than found before. They seem not to show any power law. The observed slopes at subgranular scales agree with those obtained from recent local dynamo simulations. Small differences are found for vertical component of kinetic energy that suggest that observations suffer from an instrumental effect that is not taken into account.}
% conclusions heading (optional), leave it empty if necessary
  {Local dynamo simulations quantitatively reproduce the observed magnetic energy power spectra on the scales of granulation down to the resolution limit of Hinode/SP, within the error bars inflicted by the method used and the instrumental effects replicated.}
\keywords{Sun: granulation, Sun: photosphere}

\maketitle

\section{Introduction}

Recent high-resolution observations, together with state-of-the-art MHD simulations reinforced the hypothesis, first suggested by \cite{Petrovay1993}, that the small-scale dynamo is the dominant source of magnetic field in internetwork regions. Only indirect support was offered so far. It was shown that a small-scale dynamo can be efficiently sustained in strongly stratified, compressible, and non-helical surface convection without enforced recirculation \citep{vogler07,PietarilaGraham:etal:2010}. The overall structure of the magnetic field resulting from such an action seems to agree with observations, but fall short by about a factor of 2-3 in the field strength \citep{manfred2008,Danilovic:etal:2010}. Small scale dynamo solutions that are in agreement with the field strength implied by observations require a setup that mimics a deep magnetized convection zone and account for an upward directed Poynting flux in upflow regions at the bottom boundary \citep{Rempel2014}. Additional support for a small scale dynamo comes from observations where it has been shown that the mean unsigned magnetic flux in the internetwork is not dependent on the solar cycle or location on the solar disc \citep{trujillo04,David2013}. Furthermore, the net flux imbalance in internetwork is not correlated to the surrounding network \citep{lites2011}.

Demonstrating small-scale dynamo action is difficult. If the dynamo were operating in the kinematic regime, it would be sufficient to demonstrate that the magnetic power spectrum peaks at scales smaller than the kinetic power spectrum \citep{Abramenko2011}. However, in the saturated state, such as we find in most astrophysical contexts, the power in both the magnetic field and velocity moves to larger wavelengths and the peak in the magnetic power spectrum no longer needs to be at smaller wavelengths than the peak in the kinetic energy power spectrum \citep{Moll}. Nonetheless, the power spectra remain an important diagnostic, but retrieving them is in itself a challenging task. One is not only limited by instrumental effects, but also by methods and diagnostics used to obtain information on velocity and magnetic fields. Recent studies \citep[][and references therein]{Abramenko2001,Abramenko2011,Katsukawa2012,Stenflo2012} showed that the shallow magnetic energy spectrum tends to extend towards higher wavenumbers as the spatial resolution of the instrumentation improves, but the slope remains steep at the subgranular scales. 

The largest slope value of approximately $-1$ was fitted by \cite{Katsukawa2012}, who accounted for the  Modulation Transfer Function (MTF) of the instrument. In their study, a simple deconvolution was applied, not directly to the observables, but to the derivatives - proxies for line-of-sight (LOS) velocity and magnetic flux densities.

In this paper, we perform a power spectral analysis on the same Hinode/SP \citep{Lites:etal:2001,Kosugi:etal:2007} data, but to retrieve the magnetic field and the LOS velocity, we use 2D inversions \citep{Michiel:2012}. As demonstrated by 
\cite{Michiel:2012}, the advantage of this code over the simple deconvolution is using the information contained in the full observed spectral range, simultaneously. In this way, the 2D inversions are not only able to retrieve the information up to the instrumental diffraction limit, but also to minimize the influence of noise on the retrieved highest spatial frequencies. In \cite{Danilovic:etal:2015}, we tested the code on three different simulations that give the same level of spectropolarimetric signals as the quiet Sun observations. We showed that the inversion code behaves well when a certain combination of node positions is chosen. We demonstrated that in this case, the code is able to retrieve the overall distributions of the field strength and inclination.

In this paper we concentrate on kinetic and magnetic power spectra. Again, we use the comprehensive MHD simulations to estimate how well and in what case can we recover the power spectra correctly. We also quantitatively compare our with results obtained by \cite{Katsukawa2012}.

%\begin{figure}
%\includegraphics[angle=-90,width=\linewidth,trim= 0cm 0cm 0cm 0cm,clip=true]{001000_60_maps_t0.eps}
%\caption{Results of 2D inversions applied to simulations - comparison of maps of velocity (top row), longitudinal (middle row) and transverse apparent flux density (bottom row) at $\log  \tau =0$ in Sim~2. Three columns, from left to right, show the original unsmeared maps from the snapshot, the same maps after the highest spatial frequencies are filtered out and the results from the inversions. \label{fig:sim_map}}
%\end{figure}

\begin{figure}
\includegraphics[angle=90,width=0.95\linewidth,trim= 0cm 0cm 1.5cm 0cm,clip=true]{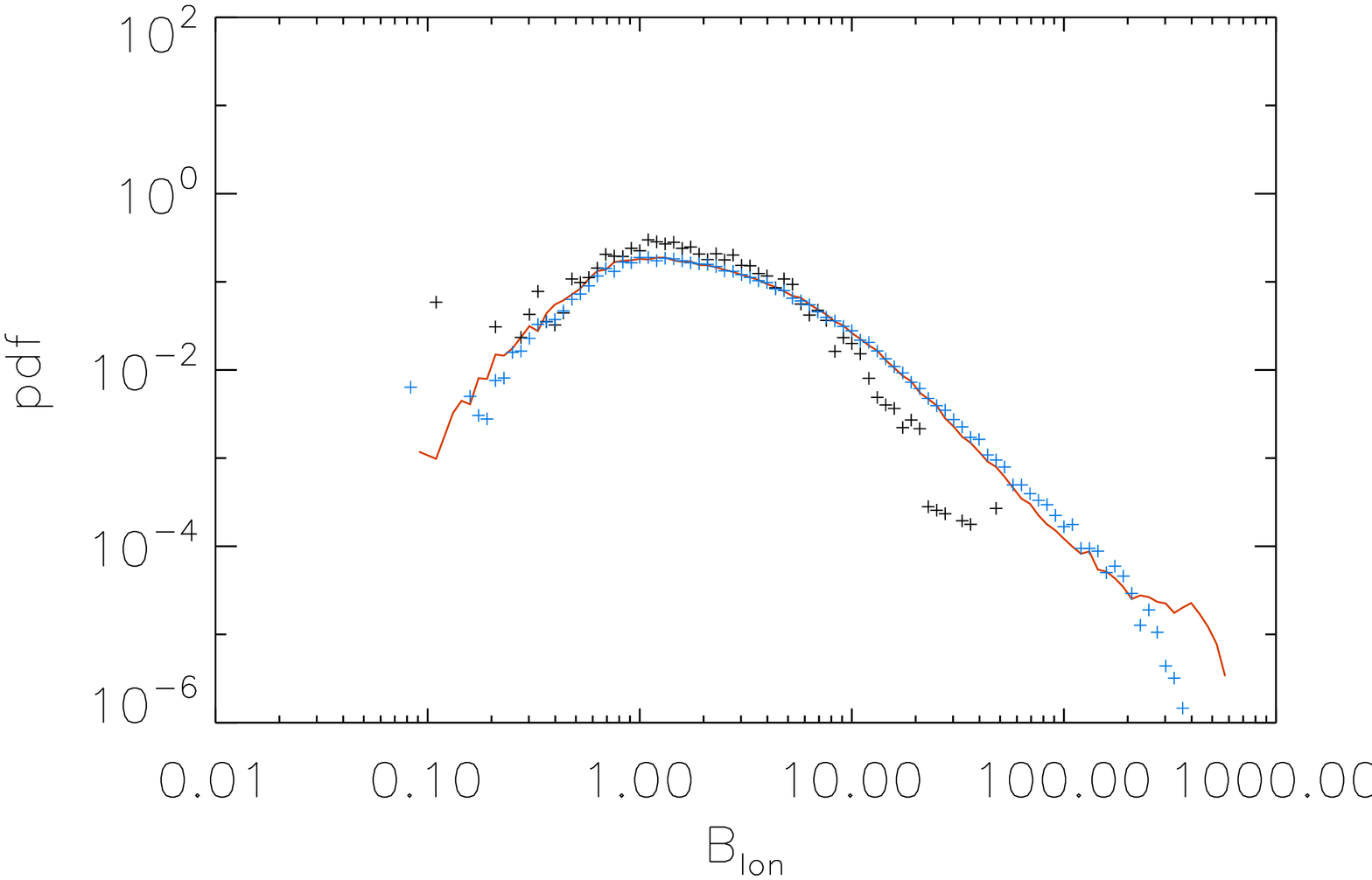}
\includegraphics[angle=90,width=0.95\linewidth,trim= 0cm 0cm 1.5cm 0cm,clip=true]{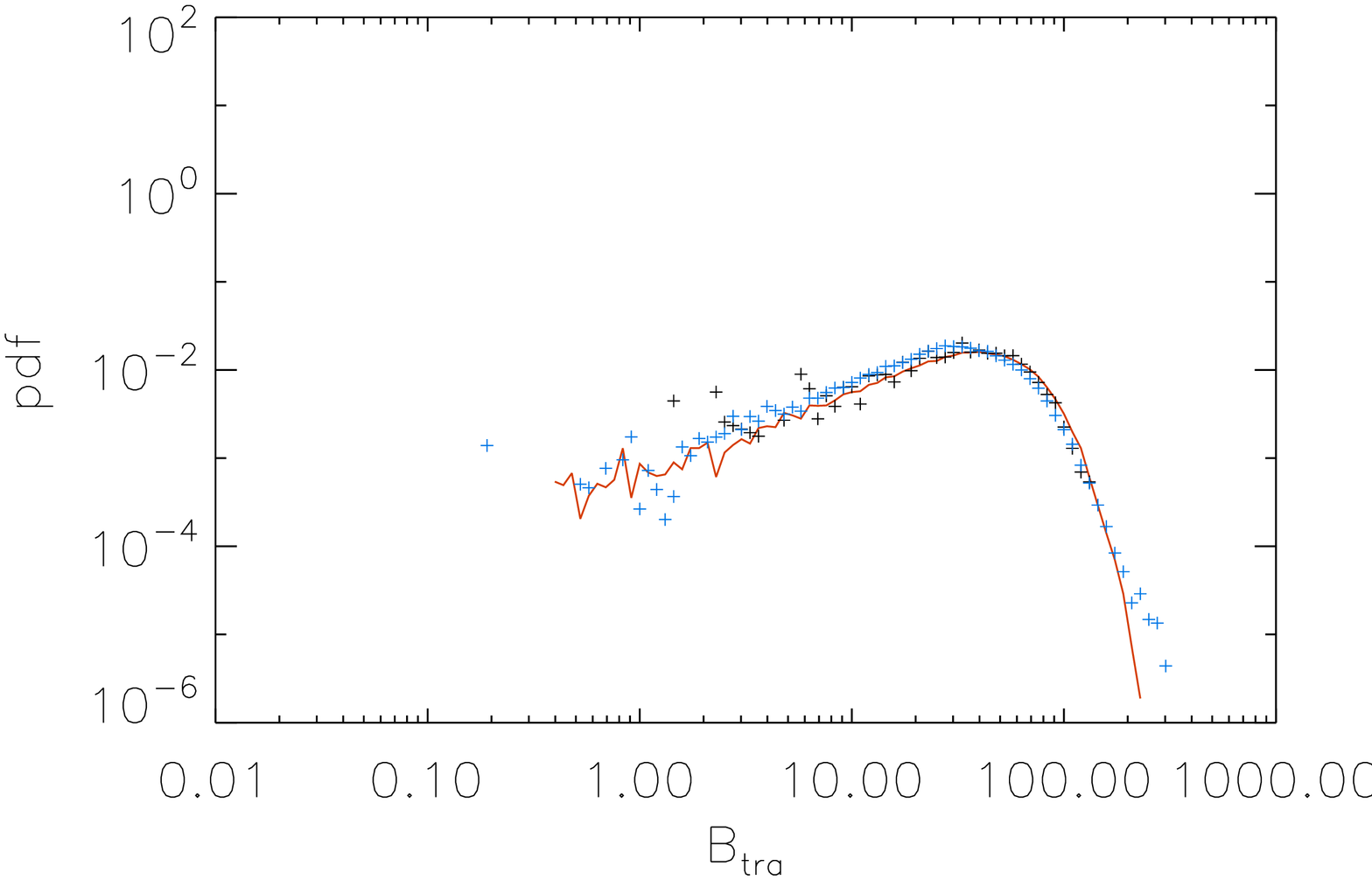}
\caption{Probability density functions (PDFs) for the longitudinal (upper panel) and transverse (lower panel) apparent magnetic flux density. PDFs from MHD simulations reduced to Hinode 
resolution (Sim1:black plus signs and Sim2: blue plus signs) are compared with the PDFs obtained from observations (red solid line). \label{fig:obs_pdf}}
\end{figure}

% trim = down right up left
\begin{figure}
\includegraphics[angle=90,width=0.95\linewidth,trim= 3.5cm 0cm 1cm 0cm,clip=true]{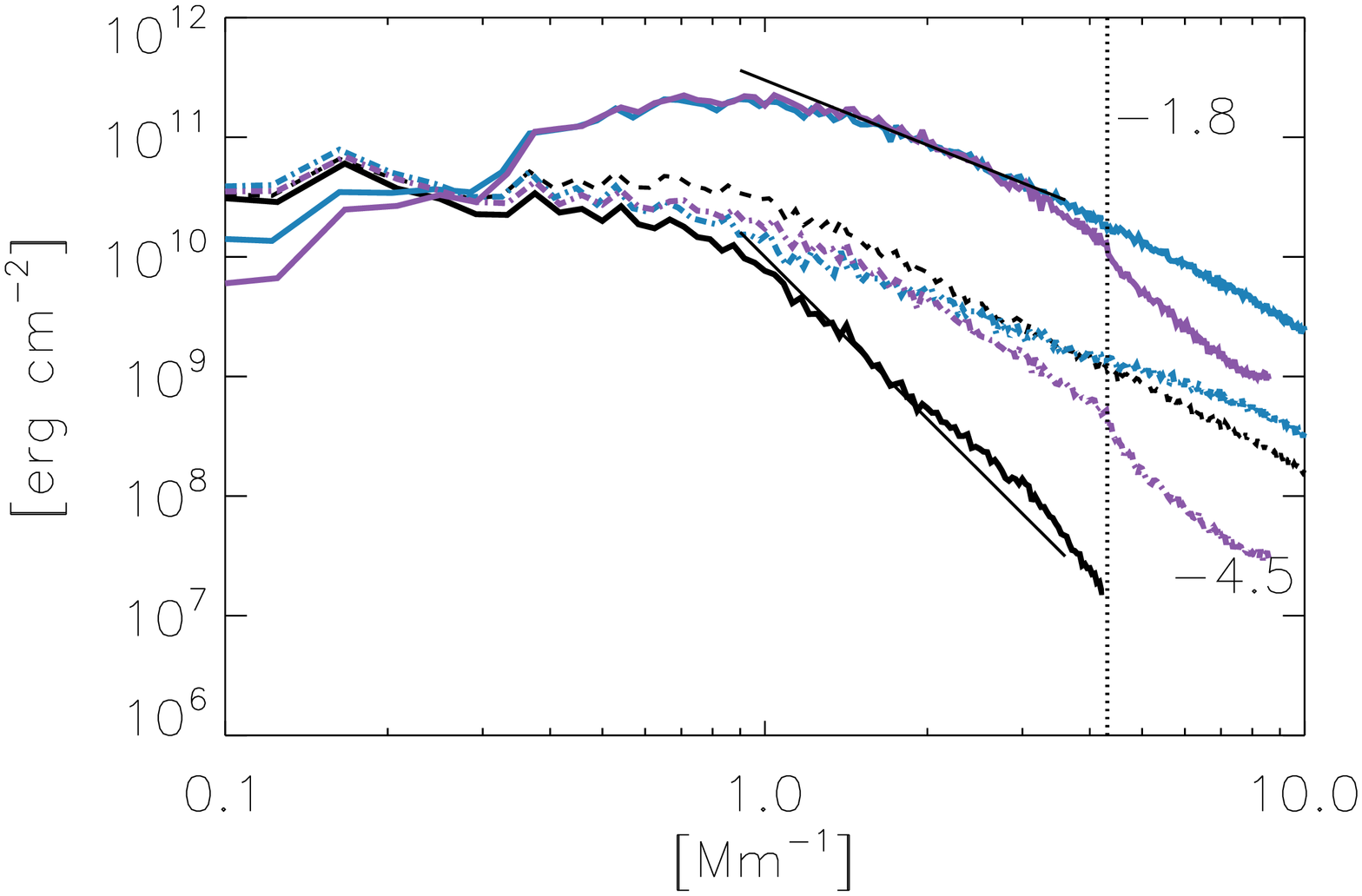}
\includegraphics[angle=90,width=0.95\linewidth,trim= 3.5cm 0cm 1cm 0cm,clip=true]{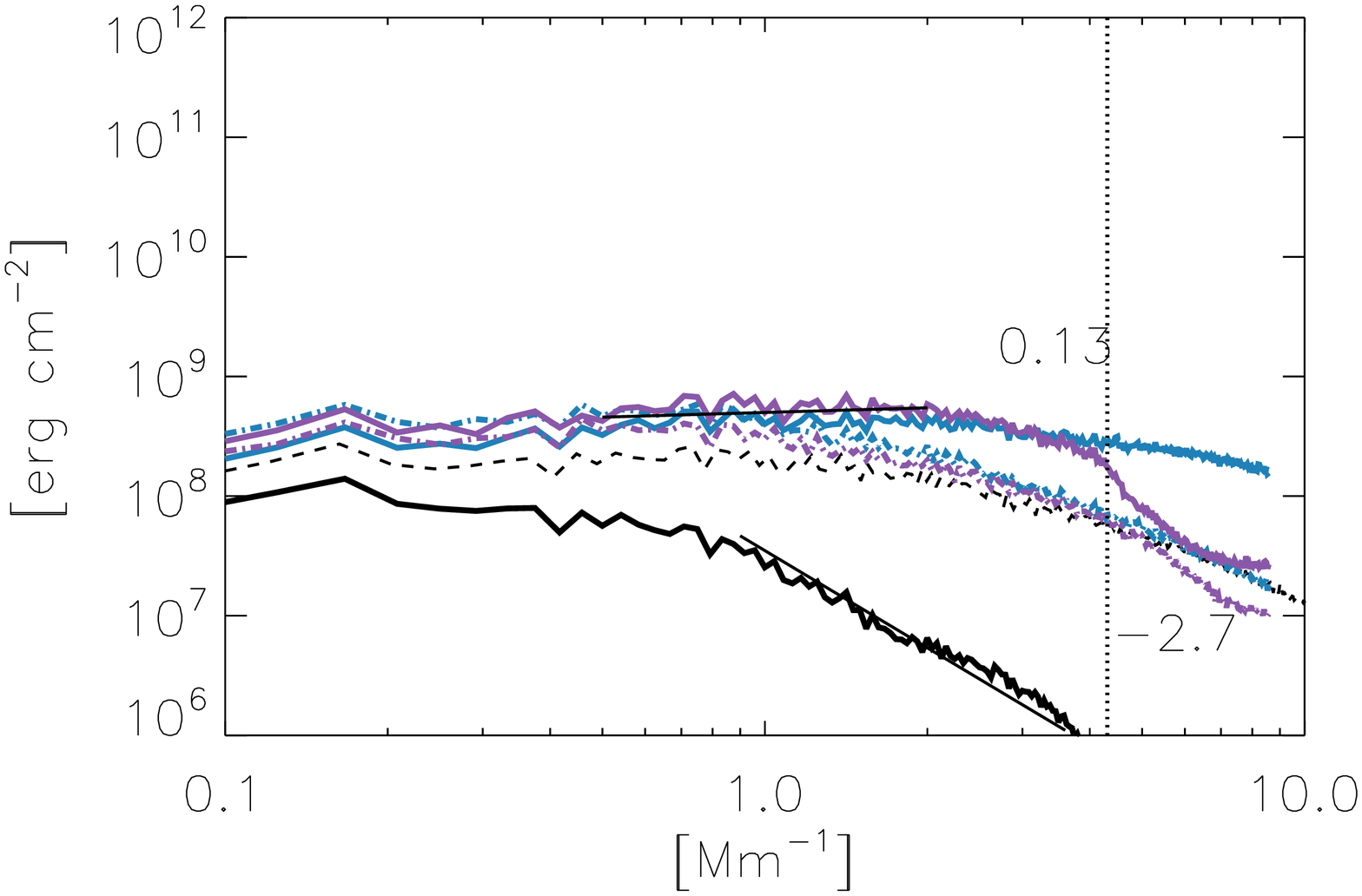}
\includegraphics[angle=90,width=0.95\linewidth,trim= 0cm 0cm 1cm 0cm,clip=true]{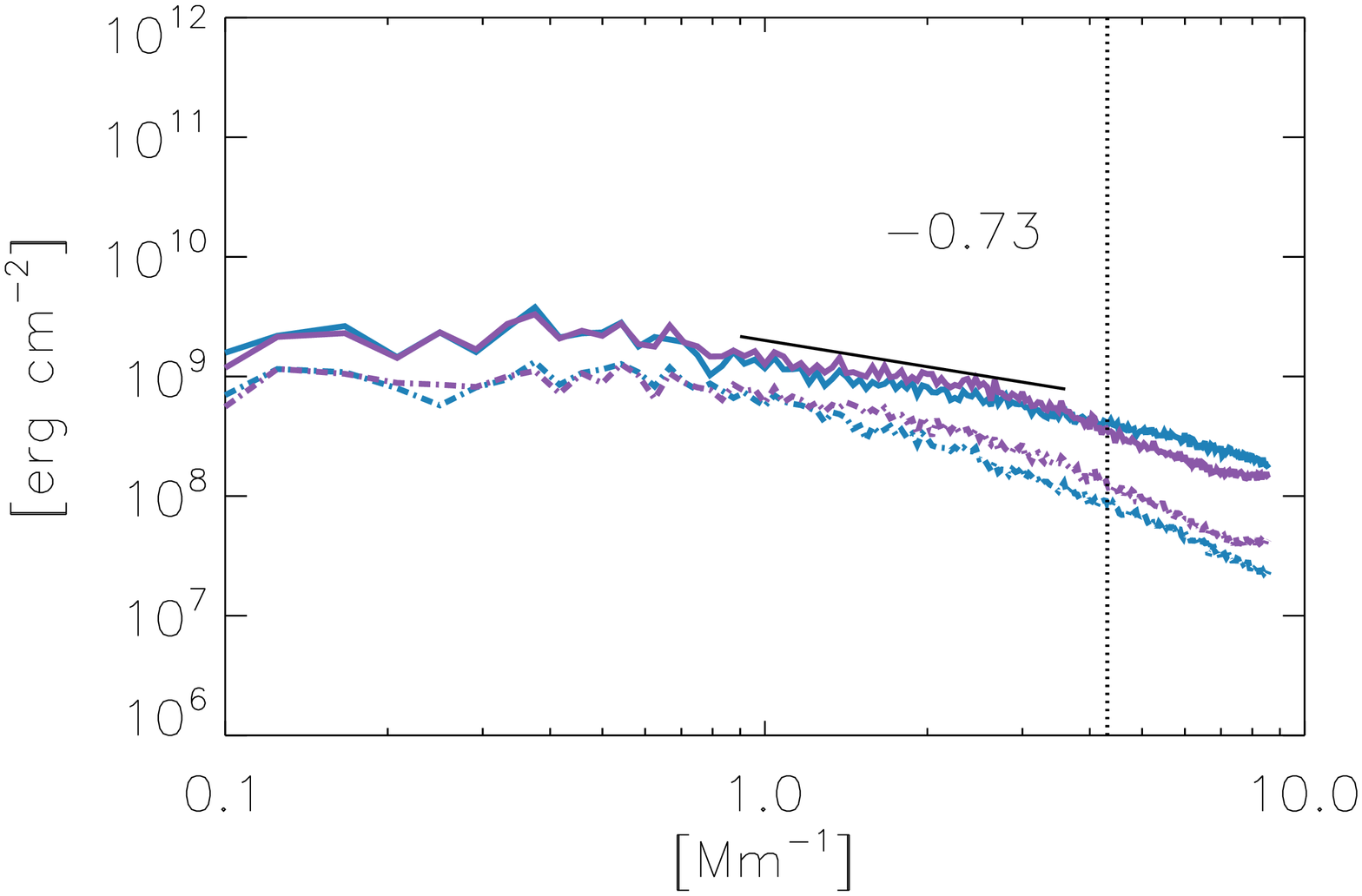}
\caption{Sim~2: Power spectra of the vertical component of kinetic (top panel) and magnetic energy (middle panel) and power spectra of the total magnetic energy (bottom panel). Blue lines mark the original spectra before any spatial smearing at different heights ($\log  \tau =0$ solid and $-2.0$ dashed). Purple lines show the result of 2D inversions at the same optical depths. Black lines are spectra of the corresponding parameters obtained from  \textit{Solarsoft} routines (see the text) before (dashed line) and after (solid line) spatial smearing. Vertical line marks the resolution limit of Hinode/SP. \label{fig:sim_power}}
\end{figure}

\begin{figure}
\includegraphics[angle=90,width=0.95\linewidth,trim= 3.5cm 0cm 1cm 0cm,clip=true]{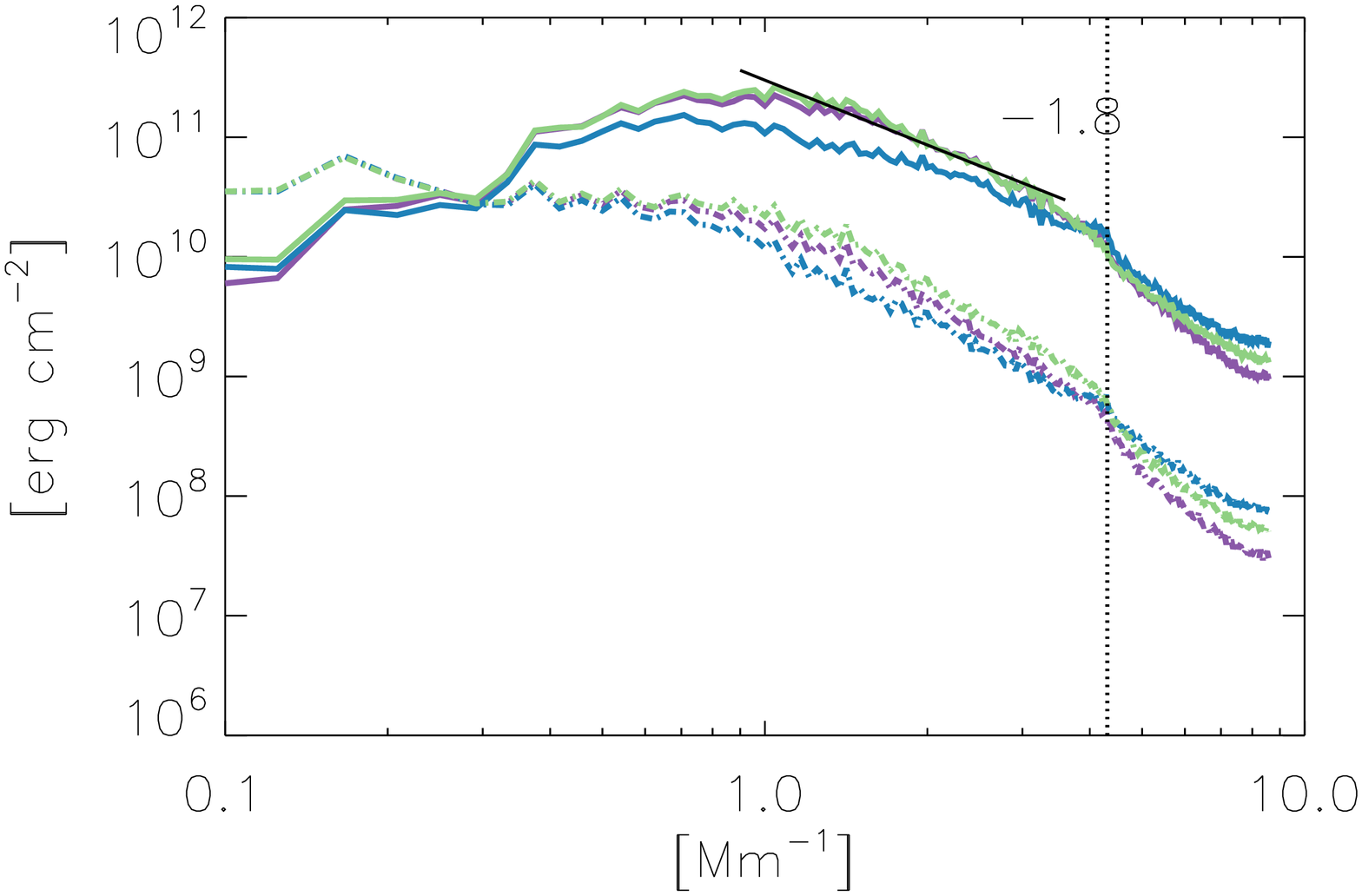}
\includegraphics[angle=90,width=0.95\linewidth,trim= 3.5cm 0cm 1cm 0cm,clip=true]{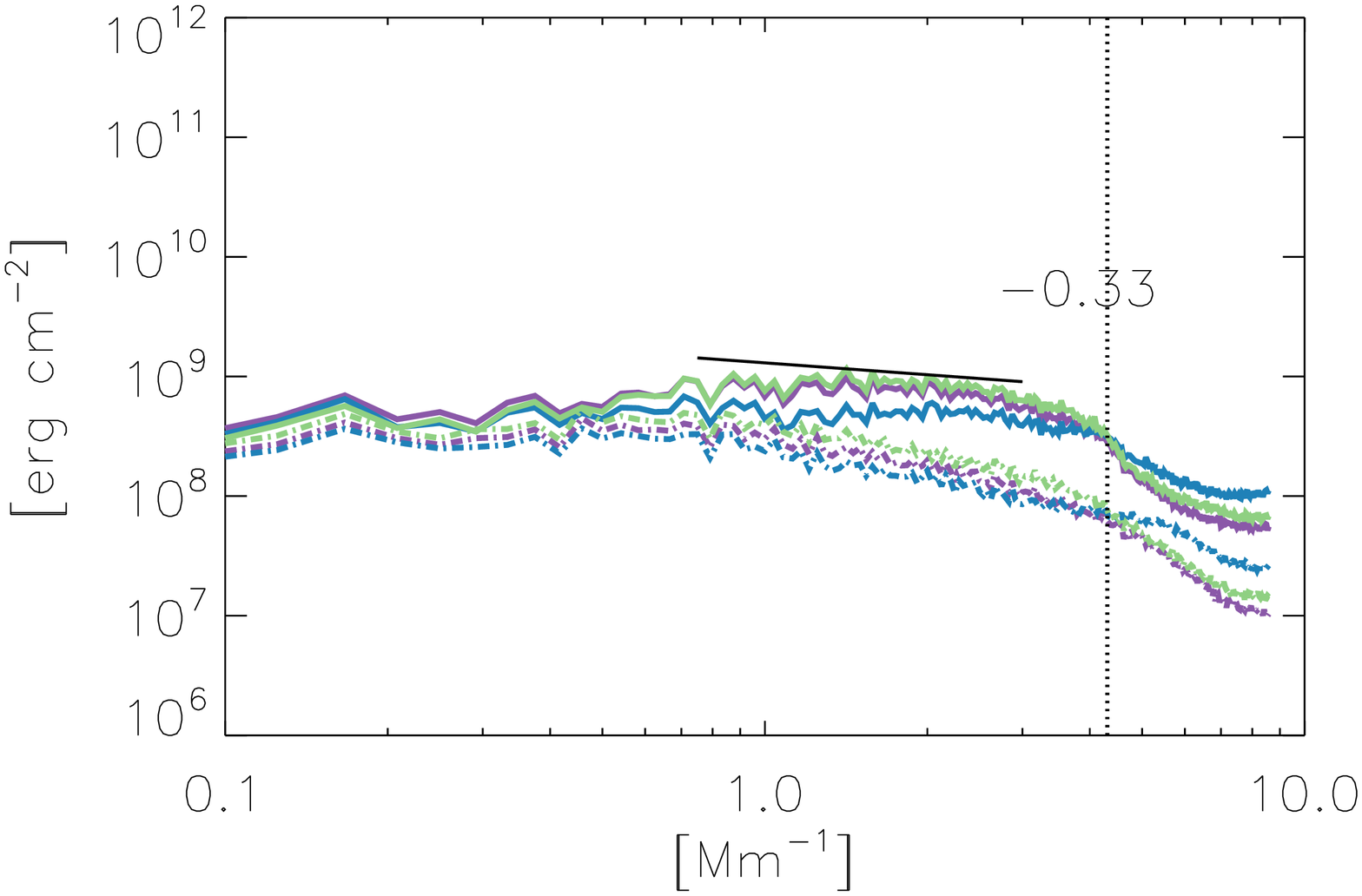}
\includegraphics[angle=90,width=0.95\linewidth,trim= 0cm 0cm 1cm 0cm,clip=true]{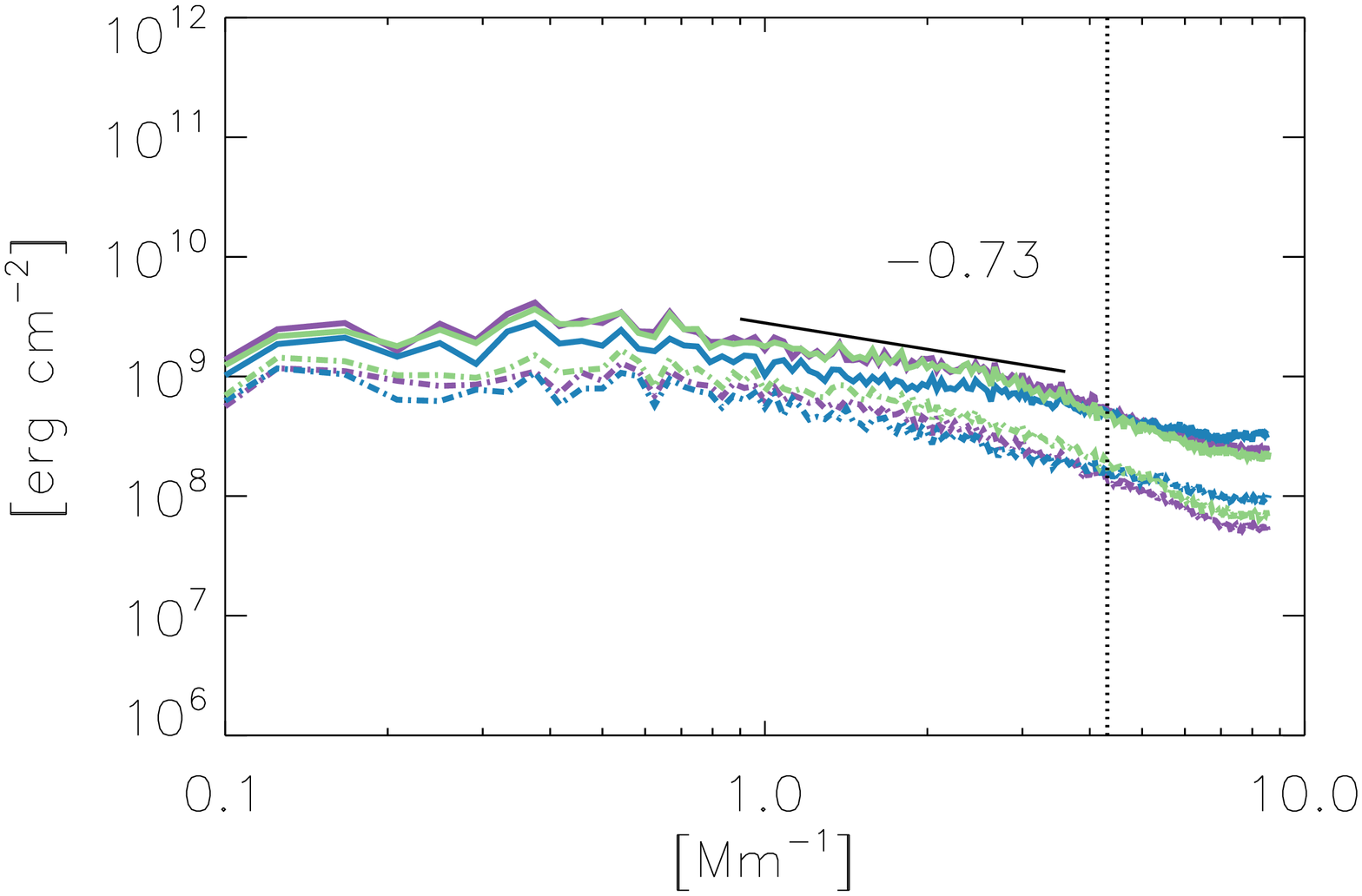}
\caption{Sim~2: Dependence of the inversion results on the PSF used. Power spectra of the vertical component of kinetic (top panel) and magnetic energy (middle panel) and power spectra of the total magnetic energy (bottom panel). Purple lines mark the spectra inverted with the correct PSF. Solid and dashed lines correspond to different heights, $\log  \tau =0$ and $-2.0$, respectively. Blue and green lines show the result when the defocus accounted for is 4 defocus steps less or more, respectively, than the one used in degradation.  \label{fig:sim_power_psfs}}
\end{figure}

\begin{figure}
\includegraphics[angle=90,width=0.95\linewidth,trim= 3.5cm 0cm 1cm 0cm,clip=true]{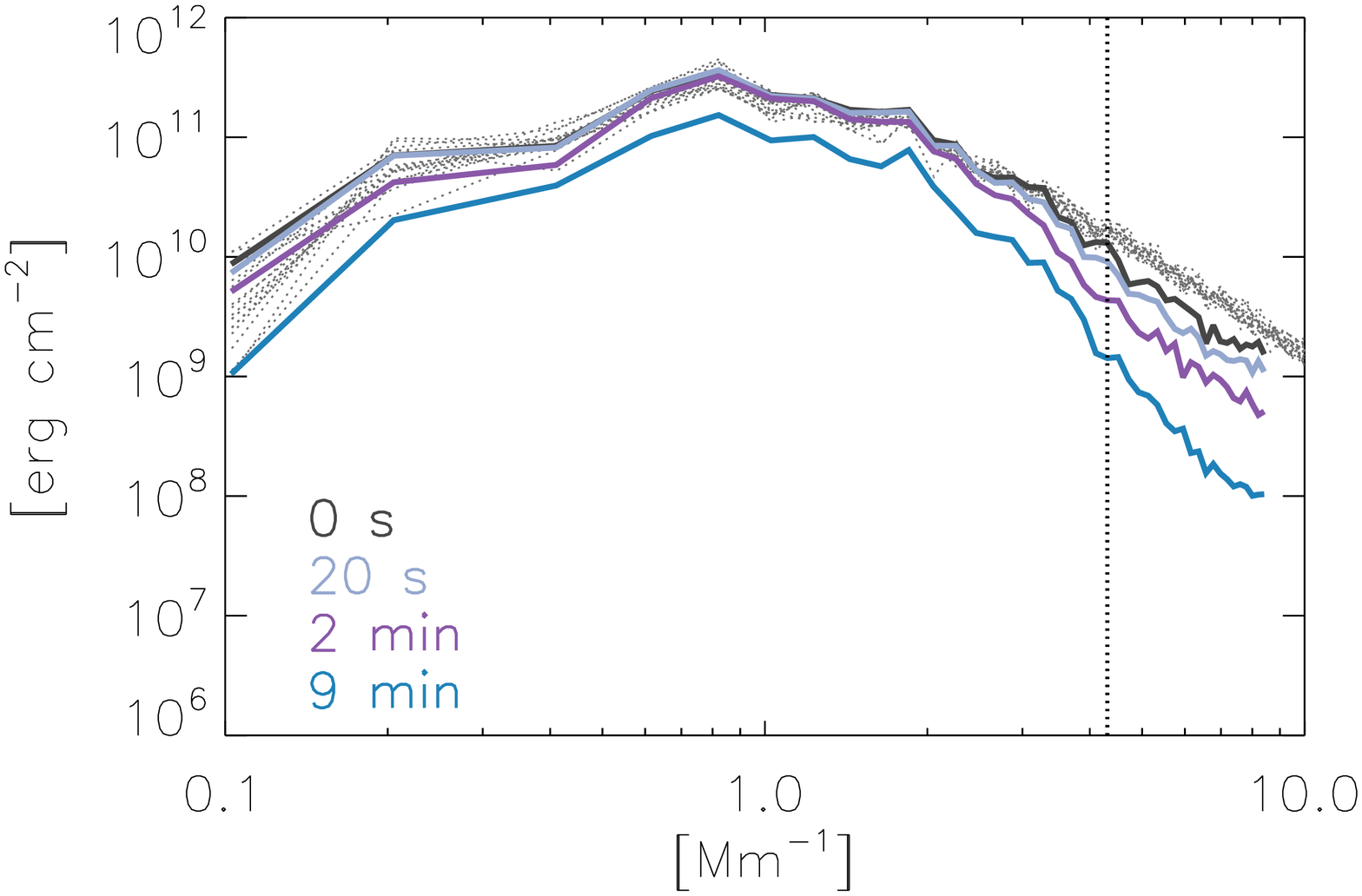}
\includegraphics[angle=90,width=0.95\linewidth,trim= 3.5cm 0cm 1cm 0cm,clip=true]{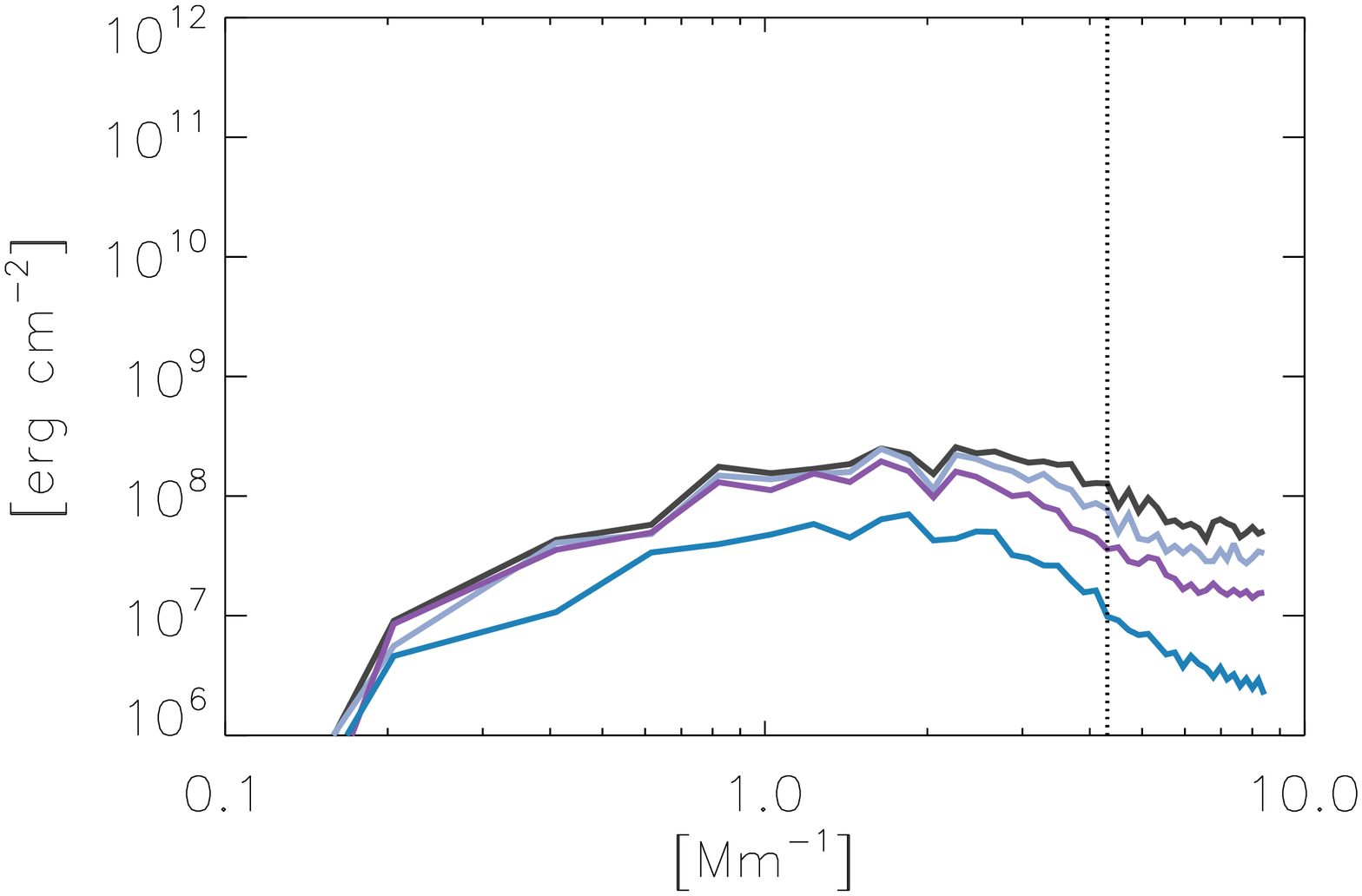}
\includegraphics[angle=90,width=0.95\linewidth,trim= 0cm 0cm 1cm 0cm,clip=true]{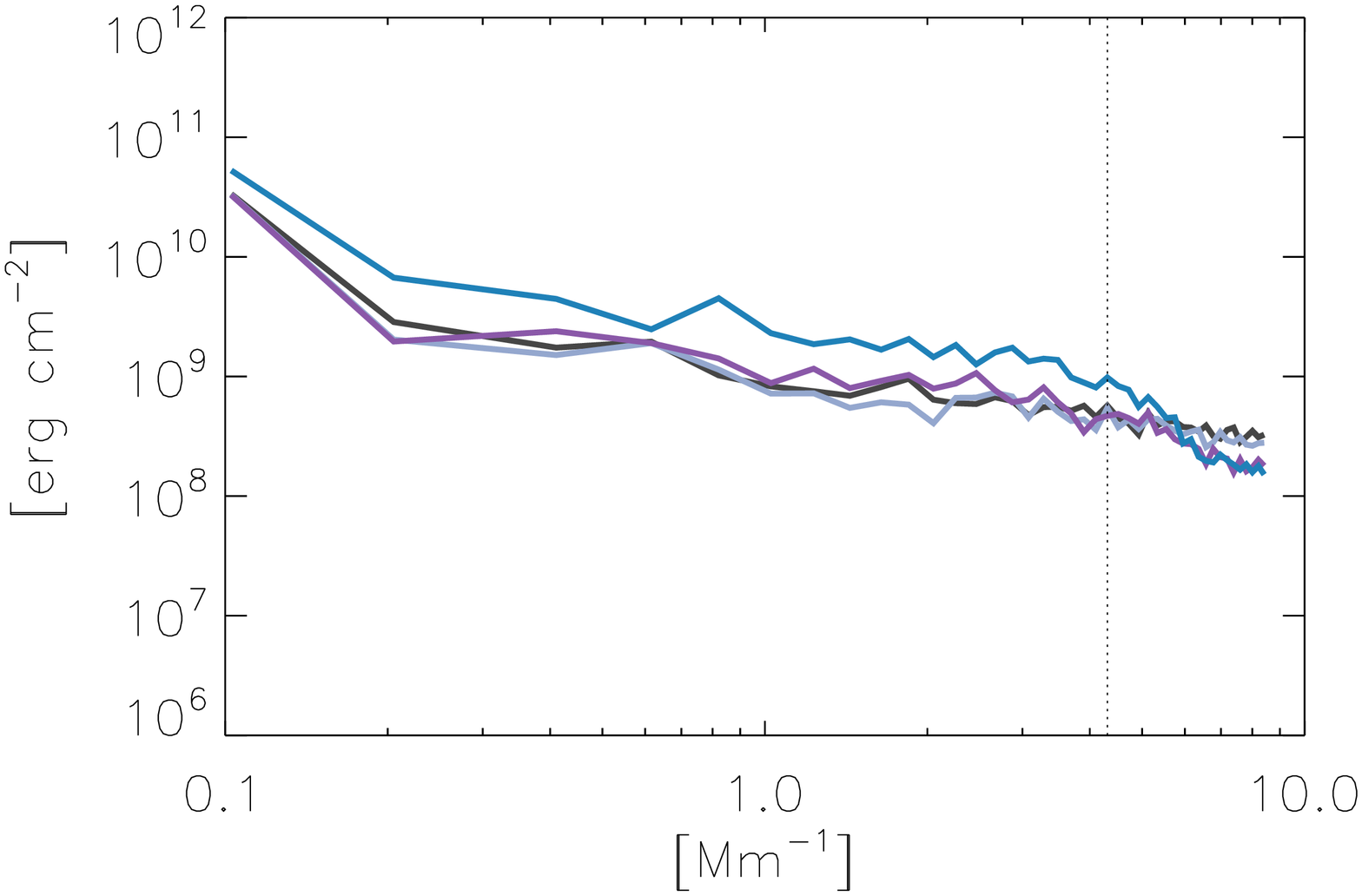}
\caption{Sim~1: How temporal averaging influences the power spectra of the vertical component of kinetic (top panel) and magnetic energy (middle panel) and power spectra of the total magnetic energy (bottom panel). Different colors mark the power spectra obtained after integration over 0~s, 20~s, 2~min and 9~min as denoted in the legend. \label{fig:sim_power_evol}}
\end{figure}

%\begin{figure}
%\includegraphics[angle=90,width=0.95\linewidth,trim= 0cm 0cm 1cm 0cm,clip=true]{vel_hist_338500_rempel.eps}
%\caption{Comparison of vertical component of velocity in old dynamo ultra simulations and new Rempels run with 16~km resolution. \label{fig:sim_power_evol}}
%\end{figure}

%\begin{figure}
%\includegraphics[angle=-90,width=\linewidth,trim= 0cm 0cm 0cm 0cm,clip=true]{rempel_obs_maps_vt1.eps}
%\caption{Observations: Field of view used in the analysis. \textit{From top to bottom:} Line-of-sight (LOS) velocity, longitudinal and transverse magnetic flux density. Parameters on the left were obtained from  \textit{solarsoft} routines (see the text). The corresponding maps on right are the results of 2D inversions.  \label{fig:obs_map}}
%\end{figure}

\begin{figure}
\includegraphics[angle=90,width=0.95\linewidth,trim= 3.5cm 0cm 1cm 0cm,clip=true]{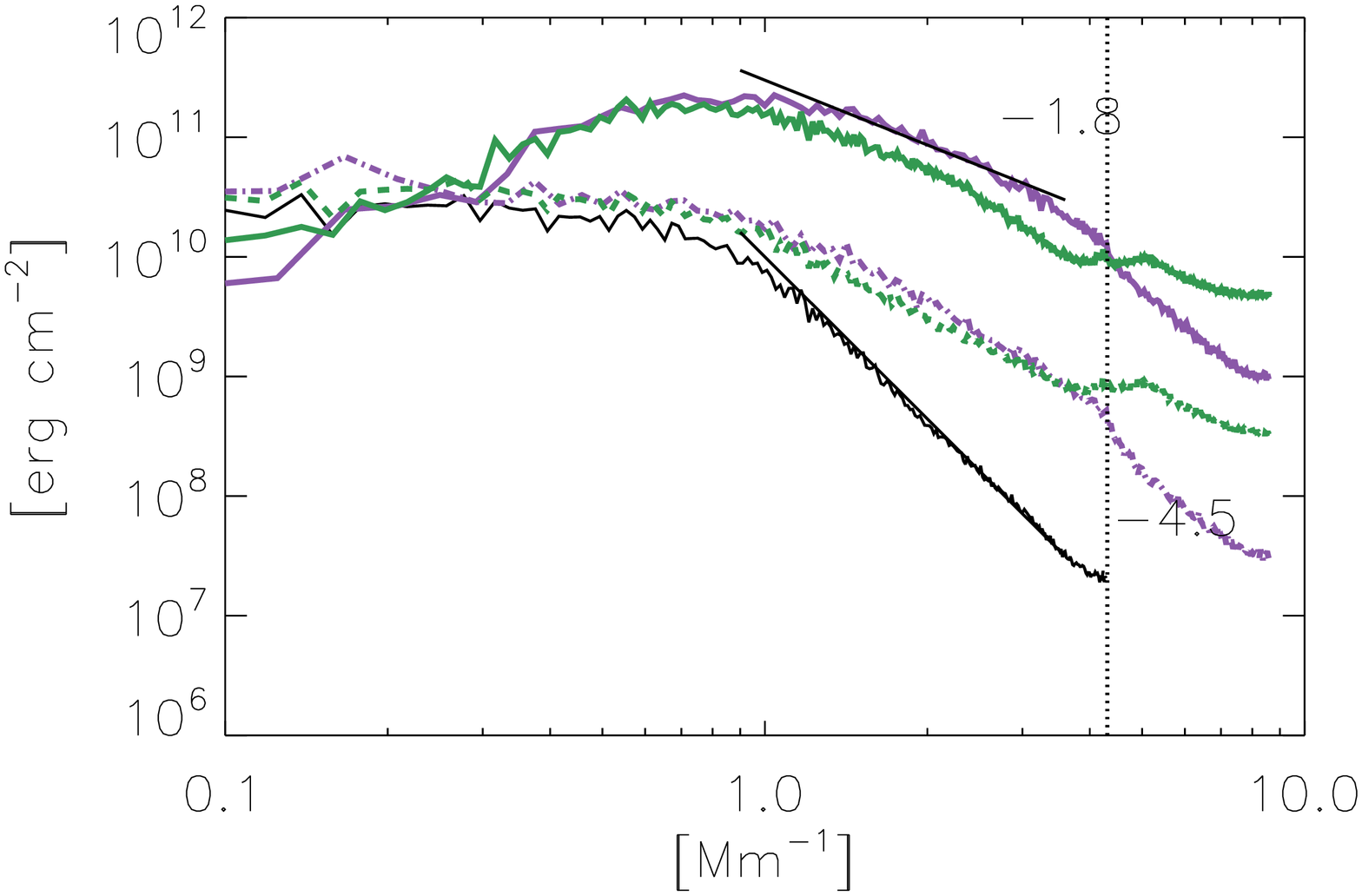}
\includegraphics[angle=90,width=0.95\linewidth,trim= 3.5cm 0cm 1cm 0cm,clip=true]{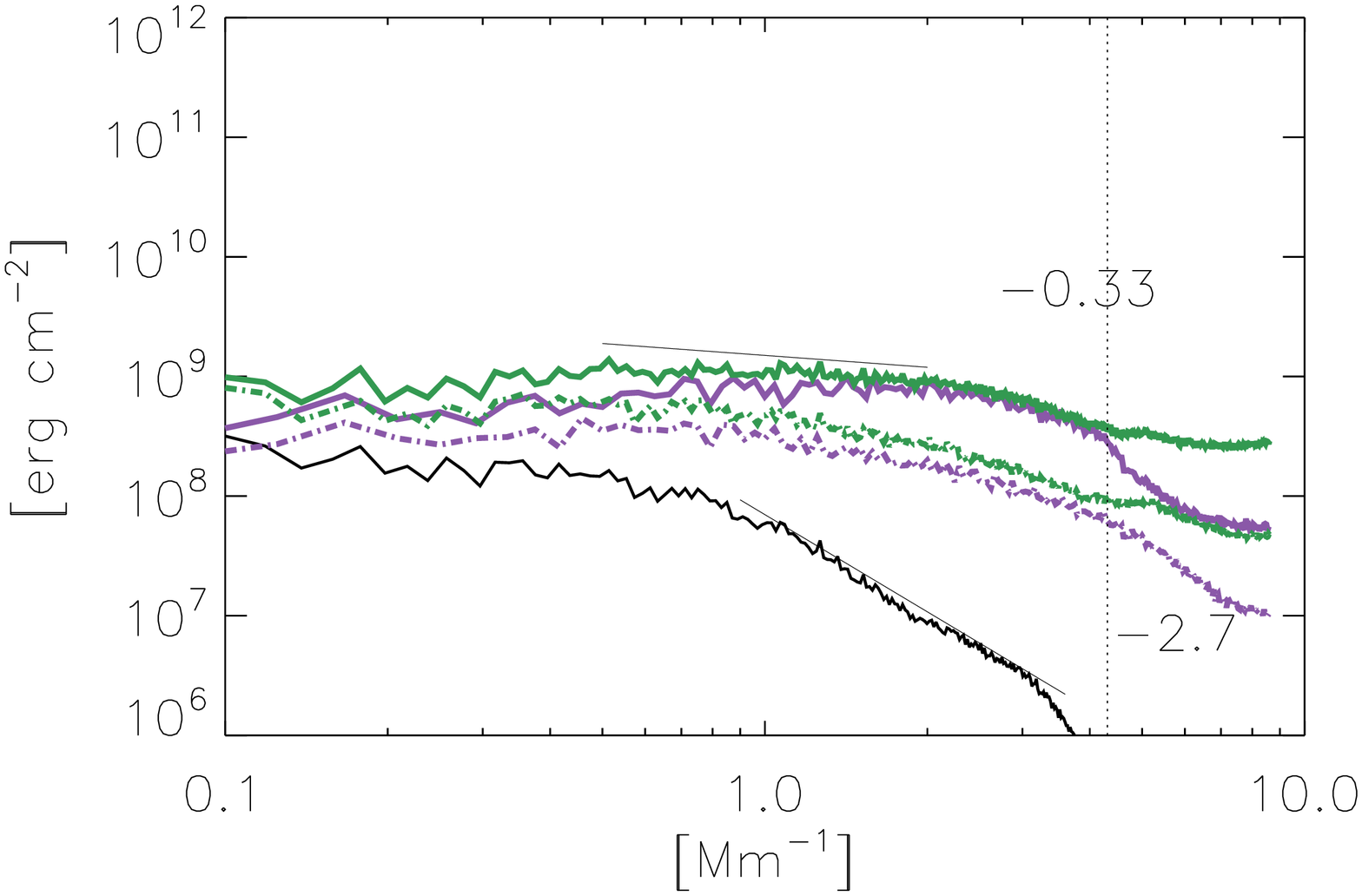}
\includegraphics[angle=90,width=0.95\linewidth,trim= 0cm 0cm 1cm 0cm,clip=true]{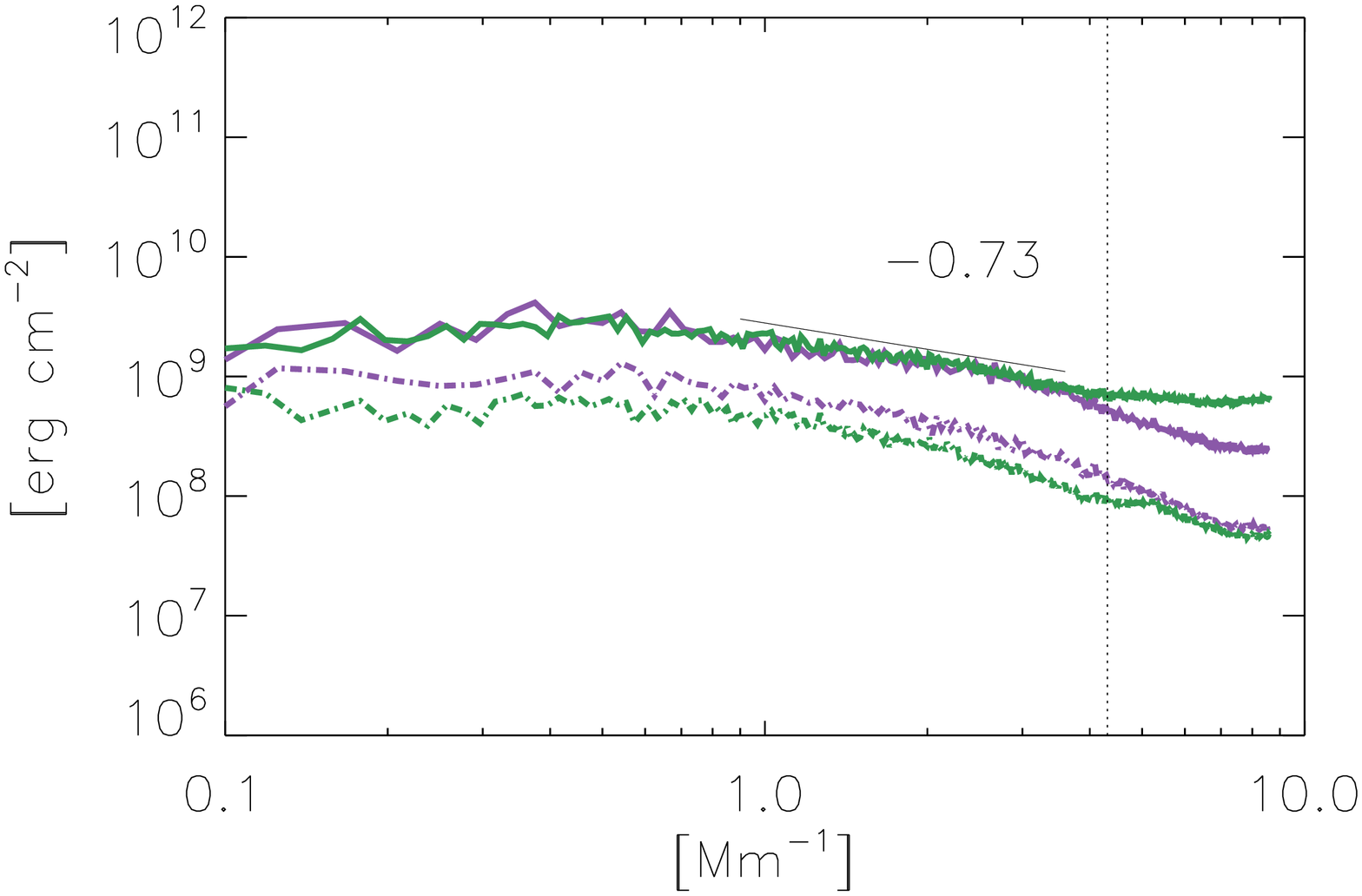}
\caption{Observations: Power spectra of the vertical component of kinetic (top panel) and magnetic energy (middle panel) and power spectra of the total magnetic energy (bottom panel). Green lines mark the results of 2D inversions at different heights ($\log  \tau =0$ solid and $-2.0$ dashed). Black lines are spectra of the corresponding parameters obtained from  \textit{Solarsoft} routines (see the text). Purple lines mark the curves obtained from Sim~2 (purple curves in Fig.~\ref{fig:sim_power}). \label{fig:obs_power}}
\end{figure}

\section{Test on MHD simulations}

We take two different local dynamo simulations produced by the MURaM code \citep{Voegler:etal:2005}. The first, here called Sim~1, is the Run~G ($R_m \sim 5200$) previously used in \cite{Danilovic:etal:2010} and \cite{PietarilaGraham:etal:2009,PietarilaGraham:etal:2010}. This run covers $4.86\times1.4\times4.86$~Mm with the resolution of 5~km. It simulates dynamo action, without any magnetic flux coming through the bottom boundary. Because of this, it reaches lower magnetic energy in the saturation phase than what is necessary to quantitatively reproduce the observations. For this reason, we multiply the field strength by a factor of 2, following the same argument as in  \cite{Danilovic:etal:2010}. We also use one snapshot from the most recent local dynamo simulation (hereafter Sim~2), that takes into account the coupling between the top layers with the bulk of the convection zone \citep{Rempel2014}. The computational domain, in this case, is $24\times7.68 \times24$~Mm, with some $1.5$~Mm above $\tau_{500}=1$, and a resolution of $16$~km in both the horizontal and vertical direction. This run uses an open bottom boundary condition that allows the presence of (small-scale) horizontal field in upflow regions in an attempt to mimic a deep magnetized convection zone (implemented through a symmetric boundary condition for all three magnetic field components). It was shown by \cite{Rempel2014} that only such a boundary condition or alternatively a closed bottom boundary with complete recirculation leads to the observationally inferred field strength of the quiet sun. In this case, the mean unsigned vertical field strength at $\tau=1$ is approximately $60$~G.

Figure~\ref{fig:obs_pdf} shows the comparison of the synthesized observables and the  Hinode/SP deep magnetogram observations.\footnote{The x-axis is chosen to be logarithmic so that distribution in the hG field range would be better visible and the results consistent with \cite{Danilovic:etal:2010}. However, the values bellow the noise level (B$^{L}_{app}=0.5$~G and B$^{T}_{app}=20$~G) should be disregarded.} The simulations are treated in the same way as in \cite{Danilovic:etal:2010}. After taking into account all instrumental effects, the longitudinal and transverse flux densities are calculated by using the \textit{Solarsoft} procedure \citep{Lites:2013}. Figure~\ref{fig:obs_pdf} shows the comparison with data set II from \cite{Danilovic:etal:2010}, where the noise level was reduced to $3\times10^{-4}$~I$_{c}$ by extending the effective exposure time. Compared to the local dynamo simulations with conservative boundary conditions (Sim~1), the snapshot of Sim~2 provides a better match to the observed magnetic flux densities. The tails of the observed distributions are better reproduced  because of the deeper lower boundary that allows formation of stronger magnetic structures on the larger spatial scales. 

To retrieve the magnetic field and velocity from simulated observations, we used the same strategy as in \cite{Danilovic:etal:2015}. The fitted model is described as a height dependent atmosphere at three nodes in optical depth, with the following free parameters: temperature, magnetic field strength, magnetic field inclination with respect to the (LOS), azimuth of the magnetic field vector, LOS velocity, and a microturbulent velocity. The nodes are placed at $\log  \tau =0, -0.8$ and $-2.0$. This proved to give recovered velocity, field strength and inclination closest to the original. 

Figure~\ref{fig:sim_power} shows the original kinetic and magnetic energy power spectra, and the corresponding spectra retrieved with a 2D inversion. Since the full velocity field cannot be obtained from Hinode/SP maps, only the part associated with the vertical component $ \frac{1}{2} \rho v_{z}^{2}$ is discussed here. A similar plot is added for the vertical component of the magnetic energy in order to compare it to the results obtained in previous studies. For reference, Fig.~\ref{fig:sim_power} shows the same quantities derived by applying the standard \textit{sp\_prep Solarsoft} routines, that obtain the LOS velocity by fitting the Fe~I~$630.15$~nm line and the apparent magnetic flux by using the approximation given in \cite{lites08}, to both unsmeared and instrumentally degraded simulated data.

 %strictly speaking, we cannot use the term kinetic energy. Because of the vertical stratification, the power spectra of vertical component of velocity shows more power on granular scale and consequently has the steeper slope on subgranular scales. Only on scales smaller than 50~km, where the velocity field becomes isotropic \citep{PietarilaGraham:etal:2010}, the two curves (power spectra of total and vertical velocity) converge. \textcolor{red}{This paragraph has to be reformulated, no?..}

Since the core of the Fe~I~$630.15$~nm line is formed fairly high, it is not a surprise that the power spectrum of the velocity obtained by fitting the core sits closer to the curve showing the power spectrum of the velocity at $\log  \tau =-2.0$ than that at optical depth unity. The curves are not strictly parallel, which could be due to variations in the formation height of the Fe~I~$630.15$~nm line core. After taking into account the instrumental spatial resolution, the slope at the subgranular scales becomes steep, with a power-law index close to $-4.5$ as found by \cite{Katsukawa2012}. 

Similarly, the longitudinal apparent magnetic flux density shows significantly reduced power at subgranular scales after spatial smearing, with a power-law index close to $-2.7$, in agreement with observations \citep{Katsukawa2012}. At the original resolution, the curve sits lower than the curve corresponding to the longitudinal component of magnetic field at $\log  \tau =-2.0$. Given that the procedure for calculating the apparent flux density takes into account the full spectral range of Hinode/SP, it is hard to specify the height that the values correspond to, without performing a detailed analysis. The power spectra suggest that the procedure seems to underestimate the field strength at all spatial scales.

The red curves in Fig.~\ref{fig:sim_power} show the result of a 2D inversion applied to a simulation snapshot. All parameters are well retrieved up to the resolution limit, but beyond the smallest spatial scales (approximately around 250~km in this case), they diverge rapidly from the original. This is where the diffraction limit is located and thus no information on spatial frequencies higher than this is contained in the data. The departure from the original curve happens a bit sooner for the vertical component of velocity at $\log  \tau =-2.0$. We ascribe this to the highly complex stratification of some of the simulated atmospheres, which cannot be always retrieved by an inversions that fits the physical quantities at only 3 nodes and/or to the low signal-to-noise ratio in the line core. 

Fitting a slope of the power spectra can be very tricky and misleading as shown by \cite{Nordlund:etal:1997}. To be on the safe side, a clear behaviour over at least of an order of magnitude for the whole time series is needed. Taken that we clearly do not have such a case, we deduce that none of the retrieved power spectra seem to follow a clear power law, but we give indices just to have some quantitative measure of the slopes. If we take the spatial scales around 1~Mm$^{-1}$, we get $-1.8$ for the LOS kinetic energy power spectrum and a positive slope of $0.13$ for the longitudinal component of the magnetic energy. Considering also the transverse component of the magnetic field increases the power at both the largest and the smallest spatial scales, and makes the slope slightly negative, around $-0.73$ for the total magnetic energy, in good agreement with the slope of the original distribution.

In \cite{Danilovic:etal:2015}, we tested how a slight change of the defocus accounted in the PSF affects the retrieved distributions of the field strength and inclination. We chose three different PSFs to estimate the impact. The first PSF assumes a defocus of 7 steps away from the optimal focus position, which brings the simulated continuum contrast to the observed value, as shown in \cite{Danilovic:etal:2008}. This PSF was also a referent PSF because it is also used to spatially smear the synthesized Stokes profiles and produce the simulated Hinode observations. The other two PSFs assumed defocus which is $\pm4$ defocus steps away from the reference, in both directions - towards the optimal focus position and away from it. When the defocus value accounted in the PSF that used in 2D inversion is changed by 4 defocus steps, the effect on the retrieved distributions of the field strength and inclination was found to be minimal. In this case, the impact was the largest in the weak end of the field strength distribution, where the field strengths were underestimated when the PSF used in the inversion is too compact. When looking at the power spectra, however, when the PSF used to invert the data is slightly wrong, the recovered power spectra diverge significantly from the original, especially for the velocity. Figure~\ref{fig:sim_power_psfs} shows the difference when the same test is applied on the power spectra. An underestimate of the width of the PSF\footnote{when the PSF used in 2D inversions assumes a defocus that is 4 defocus steps smaller than it should be.} leads to less magnetic power recovered on the granular and subgranular scales. At the same time, accounting for the wider PSF results in overestimate at the same spatial scales. The kinetic power spectrum follows the same pattern, where the sensitivity to an under-estimate of the width of the PSF seems to be larger than that to an over-estimate. 

Another potential source of systematic error is temporal averaging due to the finite exposure time, which has to be taken into account when comparing with observations. Figure~\ref{fig:sim_power_evol} shows what happens with the retrieved power spectra after temporal averaging, demonstrated on Sim~1. As in \cite{Danilovic:etal:2015}, we choose to integrate over 20~s, 2~min and 9~min. The upper panel shows the range of power coming from the quasi-periodic oscillations in the simulations. Every gray curve corresponds to the instantaneous original spectra obtained from each snapshot produced over the 10~min long run. 

For the vertical components of the kinetic and magnetic energy, the temporal averaging reduces the power and steepens the slopes. When averaging over up to 2~min, the differences in the power spectra are visible only on subgranular scales. Continued averaging in time tends to shift the curves downwards on all spatial scales, similar to the results of \cite{Kitiashvili2012} for horizontal velocities. The opposite effect is visible in power spectra of magnetic energy. Increase in power comes from the horizontal component of magnetic field, whose contribution grows with exposure time, as found by \cite{Danilovic:etal:2015}.

\section{Power spectra of Internetwork}

We take the scan used in various studies starting with \citep{lites08} and \citep{david07a,david07b}. It is obtained in normal mode, with an exposure time of 4.8~s, on March 10th 2007, between 11:37 and 14:37 UT. Since we are mainly interested in small scales, we limit our study to a small part of the whole scan, $70\arcsec \times 70 \arcsec$ wide. The power spectra of the parameters retrieved with a 2D inversion are shown in Fig.~\ref{fig:obs_power}. Overplotted are also power spectra calculated from simulations. The parameters derived with the \textit{Solarsoft} routines, applied to observations, show the same slopes as obtained by \cite{Katsukawa2012}.  

All the curves retrieved from observations have similar slopes as those from simulations. Although it appears to have the same slope, the observed velocity power spectrum considerably differs from the simulated one at granular and subgranular scales. The difference is slightly smaller for an optical depth of $0.01$. The most obvious explanation for this are the presence of p-mode oscillations in the observations that cannot be easily filtered out from the scanned maps. Fig.~\ref{fig:sim_power_evol}, on the other hand shows that periodic osculations in the simulations result in a much smaller scatter of the kinetic energy power spectra, hence the difference we see in the observed and simulated power might be too big to be explained only with p-mode oscillations. Additionally, the observed velocity power spectrum shows a strange plateau around the resolution limit. Given that the same feature is somewhat less visible but still present also in the magnetic spectra, we are prone to conclude that this could be due to some instrumental effect that we did not take into account, e.g. JPEG compression or some other source of noise not introduced in the simulations. Furthermore, the reduced power in the observed velocity power spectrum on the granular and subgranular scales might be due to a small error in the PSF used in the inversion. As demonstrated in Fig.~\ref{fig:sim_power_psfs}, an over- or underestimate of the width of the PSF can result in a kinetic power spectrum with the observed slope.

Another, but smaller difference can be seen in the power spectrum of the vertical component of magnetic field. The simulated and observed power spectra seem to diverge for spatial scales larger than the granular scale. The observed spectrum shows more power at these scales, which changes the slope from slightly positive to slightly negative. The difference comes from the strong network visible in the observed maps, but not present in simulations. \cite{Rempel2014} showed that increasing the simulation domain size (both vertically and horizontally) leads to an increase of power on scales larger than granulation.

\section{Conclusions}

Applying a 2D inversion code to spectra synthesized from MHD simulated snapshots showed that all atmospheric parameters can be retrieved reliably up to the diffraction limit of the telescope when all instrumental effects are taken into account properly. The power spectra are recovered to much smaller spatial scales than with any other method used before, without being affected significantly by the limited spatial resolution. Although we cannot claim that any of the spectra follow a power-law, we find much gentle slopes at subgranular scales than previous studies. The observed magnetic power spectra follow closely the power spectra obtained from the most recent local dynamo runs, however, a mismatch of the observed and simulated kinetic power spectra was still observed. The inherent sensitivity of this quantity to the instrumental properties suggests that perhaps some inaccuracies in the instrumental properties still remain. Although state-of-the-art simulations show that the effect of small scale dynamo action has its peak at the scales comparable to the resolution limit of Hinode, looking at smaller scales is of course still desirable, as it is having an optical system whose properties are well known.

 \begin{acknowledgements}
  The National Center for Atmospheric Research (NCAR) is sponsored by the National Science Foundation. We would like to acknowledge high-performance computing support from Yellowstone (http://n2t.net/ark:/85065/d7wd3xhc) provided by NCAR's Computational and Information Systems Laboratory, sponsored by the National Science Foundation. Hinode is a Japanese mission developed and launched by ISAS/JAXA, with NAOJ as domestic partner and NASA and STFC (UK) as international partners. It is operated by these agencies in co-operation with ESA and NSC (Norway). 

  \end{acknowledgements}

\end{document}